\documentclass[twocolumn,aps,prd]{revtex4-1}
\pdfoutput=1

\usepackage[pdftex]{graphicx}
\usepackage[unicode=true, bookmarks=false,bookmarksnumbered=true,bookmarksopen=true, breaklinks=false,backref=false,pagebackref=false, colorlinks=true]{hyperref}
\hypersetup{pdftitle={ },pdfauthor={ },linkcolor=black,citecolor=black, urlcolor=black, filecolor=black,pdfpagelayout=OneColumn,pdfnewwindow=true,pdfstartview=XYZ, plainpages=false}
\usepackage{rotating}
\usepackage{color}
\usepackage{amsmath}
\usepackage{braket}
\usepackage[T1]{fontenc}
\usepackage{url}
\usepackage[english]{babel}
\usepackage{dcolumn}
\usepackage{bm}
\usepackage{subfig}
\usepackage{float}
\usepackage{amsmath}
\usepackage{amssymb}
\usepackage[english]{babel}

\usepackage{cleveref}
\crefname{equation}{Eq.}{Eqs.}
\Crefname{equation}{Eq.}{Eqs.}
\crefname{table}{Table}{Tables}
\Crefname{table}{Table}{Tables}
\crefname{figure}{Fig.}{Figs.}
\Crefname{figure}{Fig.}{Figs.}
\crefname{algorithm}{Alg.}{Algs.}
\Crefname{algorithm}{Alg.}{Algs.}
\crefname{section}{Section}{Sections}
\Crefname{section}{Section}{Sections}

\usepackage{multirow}
\usepackage{verbatim}

\usepackage{algpseudocode}
\usepackage{algcompatible}
\usepackage[floatrow]{trivfloat}
\trivfloat{algorithm}

\newcommand{\Tone}{\mathbb{T}}
\newcommand{\Ttwo}{\mathbb{T}^{(2)}}
\newcommand{\Tfour}{\mathbb{T}^{(4)}}

\newcommand{\TOtwo}{\mathbb{TO}^{(2)}}
\newcommand{\TOfour}{\mathbb{TO}^{(4)}}

\begin{document}

\title{Flux tube widening in compact $U(1)$ lattice gauge theory
\\
computed at $T < T_c$ with the multilevel method and GPUs.}

\author{Andr\'e Amado}
\email{andre.c.amado@ist.utl.pt}
\author{Nuno Cardoso}
\email{nuno.cardoso@ist.utl.pt}
\author{Pedro Bicudo}
\email{bicudo@ist.utl.pt}
\affiliation{CFTP, Departamento de F\'{i}sica, Instituto Superior T\'{e}cnico, Universidade T\'{e}cnica de Lisboa, Av. Rovisco Pais, 1049-001 Lisbon, Portugal}

\begin{abstract}
We utilize Polyakov loop correlations to study d=3+1 compact $U(1)$ flux tubes and the static electron-positron potential in lattice gauge theory. With the plaquette field operator, in U(1) lattice gauge theory, we probe directly the components of the electric and magnetic fields.
In order to improve the signal-to-noise ratio in the confinement phase, we apply the L\"uscher-Weiss multilevel algorithm.
Our code is written in CUDA, and we run it in NVIDIA FERMI generation GPUs, in order to achieve the necessary efficiency for our computations. We measure in detail the quantum widening of the flux tube, as a function of the intercharge  distance and at different finite temperatures $T<T_c$. Our results are compatible with the Effective String Theory.
\end{abstract}

\maketitle


\section{Introduction\label{sec:introduction}}


\subsection{Confinement and flux tubes}

Lattice Quantum Field Theory opens the way for a deeper understanding of the details of confinement, which remains a central problem of
strong interactions. In the context of this formalism we study the static-antistatic interaction and flux tube formation, providing data to be compared with theoretical models of confinement. 

A very interesting model for flux tubes is Effective String Theory, where the flux tube intrinsic width is assumed to be vanishing.
The Effective String Theory shows that many properties of confining gauge theories are universal, independent of the exact gauge group describing the interaction. One of the earlier examples of this is the derivation by L\"uscher of the form of the potential for an effective string \cite{Luscher:1980ac}
\begin{align}
	V(r) = V_0 + \sigma r - \frac{\gamma}{r} + \mathcal{O}\left(1/r^3\right)
\label{eq:potential}
\end{align}
where $\gamma = \frac{(d-2)\pi}{24}$, known as L\"uscher term, is independent of the gauge group ruling the interaction depending only on the dimension $d$ of spacetime.
Effective String Theory has also predicted how the string tension $\sigma$ and the width $w$ of the confining string should depend on  temperature.  Even with a vanishing intrinsic width, quantum fluctuations force the string to have a finite width $w$ inasmuch as the zero mode of a quantum harmonic oscillator. This subject was introduced by L\"uscher, Munster and Weisz in 1981 \cite{Luscher:1980iy}, defining the concept of flux tube widening, showing the width $w$ diverges logarithmically in chromoelectric flux tubes.

In this work, we aim to study numerically, in U(1) lattice gauge theory, the flux tubes, their different logarithmic and linear behaviors respectively of zero and finite temperatures, and to compare our numerical results with the effective string theory predictions.


\begin{figure*}[t!]
\begin{centering}
    \subfloat[In the confining phase.]{
\begin{centering}
    \includegraphics[width=0.45\textwidth]{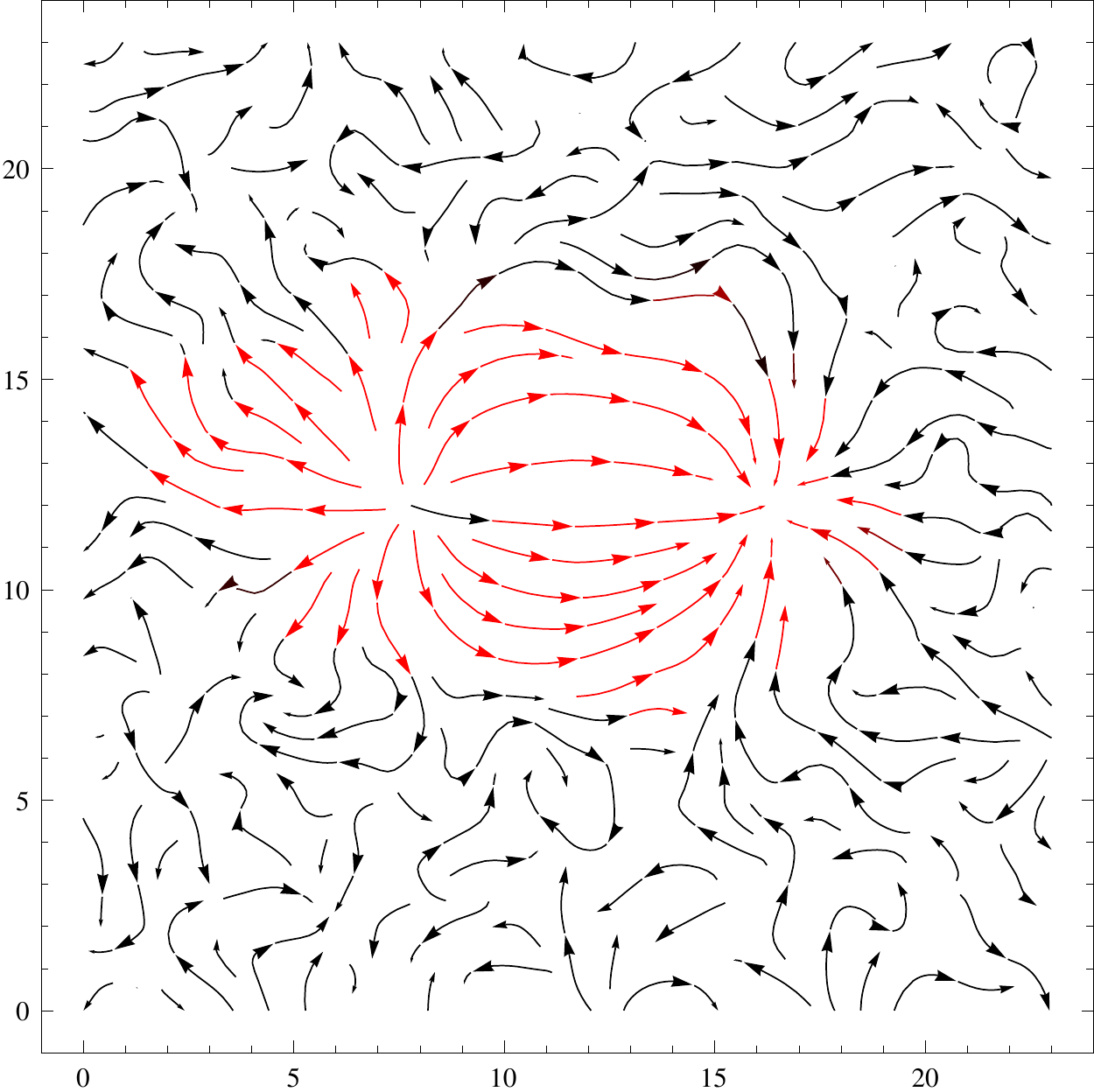}
\par\end{centering}}
    \subfloat[In the Coulomb phase.]{
\begin{centering}
    \includegraphics[width=0.45\textwidth]{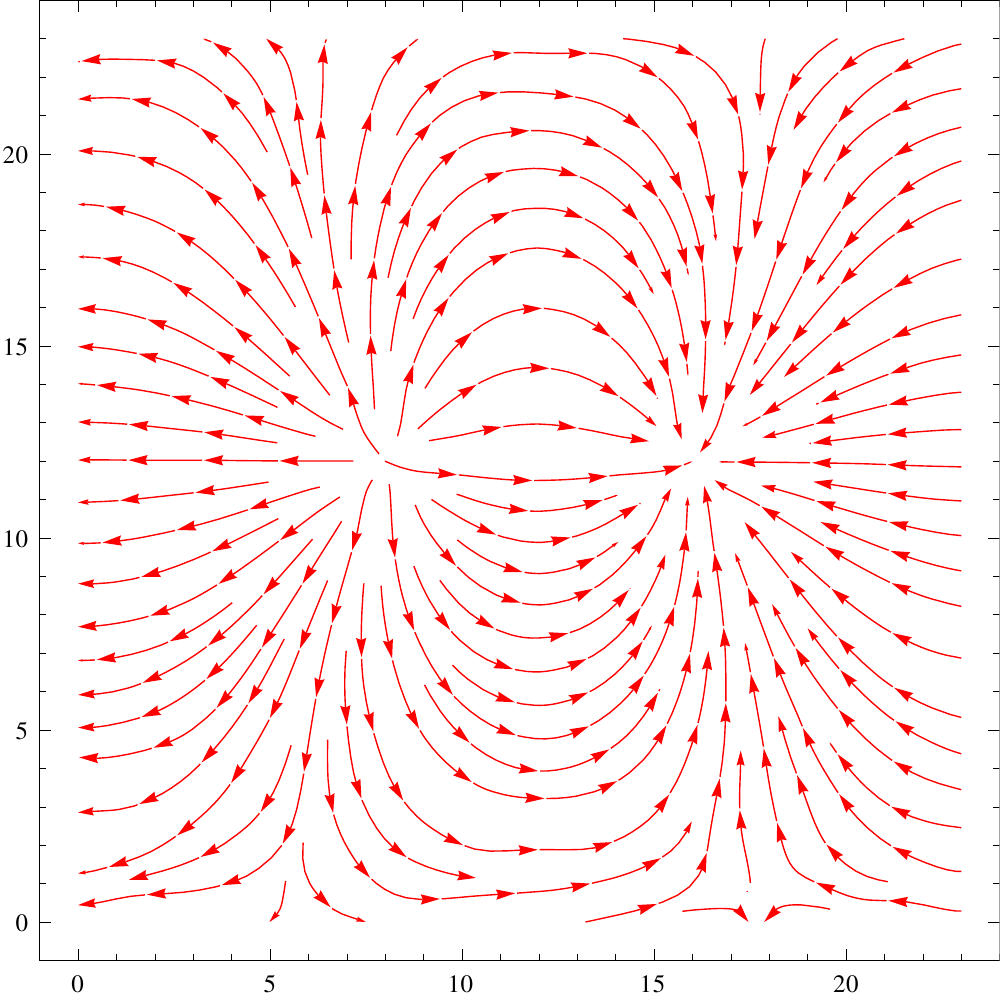}
\par\end{centering}}
\par\end{centering}
\caption{(color online) Electric field lines in the plane of the charges. The charge separation is $r=8a$. In (a), confining phase, the field intensity is only significant in to the flux tube between the charges, outside the flux tube the field is vanishing and fluctuates randomly. In (b), Coulomb phase, the field extends, with a power law decay, to the whole lattice.}
\label{fig:field_charges_plane}
\end{figure*}


\subsection{Widening of in effective string theories
\label{sub:width}}

The string width can be defined in different ways. We follow the definition used in \cite{Gliozzi:2010zv} which is
\begin{align}
w^2(r/2) = \langle h^2(r,x_\perp,t) \rangle\ ,
\end{align}
where $h(r,x_\perp,t)$ is the string representing the flux tube and the average is evaluated in the mediator plane of the charges. 
$r$ is the intercharge distance, and $x_\perp$ is the cylindrical coordinate perpendicular to the intercharge axis.

Effective string theory results for the width of flux tubes appearing in confining gauge theories have been determined in the zero and finite temperature regimes, \cite{Gliozzi:2010jh,Meyer:2010tw}, as
\begin{itemize}
\item at zero temperature 
\begin{equation} 
w^2_{lo}(r/2) = \frac{d-2}{2 \pi \sigma} \log (\frac{r}{\rho_0})
\label{eq:width_zero_temp}
\end{equation}

\item at finite temperature
\begin{equation} 
w^2_{lo}(r/2)= \frac{d-2}{2 \pi \sigma} \log \frac{1}{4 \, T \, \rho_0} + \frac{d-2}{4  \sigma} \, T \,  r + O(\mbox{e}^{-2 \pi \, T \, r})
\label{eq:width_finite_temp}
\end{equation}
\end{itemize}
where $d$ is the dimension of space-time and $\rho_0$ is some distance scale to be fitted.
These results have been recently generalized by Gliozzi et al., \cite{Gliozzi:2010zt,Gliozzi:2010zv}, who obtained the exact expression at  2-loop order, incorporating the temperature dependence. Analytic results also exist for more complex systems, such as boundstates of $k$-strings \cite{Armoni:2008sy}.
Since we have finite error bars, to compare with our lattice QCD results, the tree level result  \cite{Gliozzi:2010zt,Gliozzi:2010zv} is sufficient,
\begin{align}
w^2_{lo}(r/2) = \frac{d-2}{2 \pi \sigma} \log \left(\frac{r}{\rho_0}\right) + \frac{d-2}{\pi \sigma} \log \left(\frac{\eta(2 i u)}{\eta^2(i u)}\right)
\label{eq:width_tree_level}
\end{align}
where $\eta$ is Dedekind $\eta$ function  \cite{Gliozzi:2010zt,Gliozzi:2010zv} and $u = 1 /(2 \, T \, r)$ incorporates the temperature dependence. \cref{eq:width_tree_level} constitutes the ansatz for our fits to the widths we compute in $U(1)$ lattice gauge theory.


\subsection{ Flux tube widening in Lattice gauge theory}

Lattice gauge theory is actively progressing in checking the widening predictions of Effective String Theory in Lattice gauge theory.  Different gauge groups have been used. 

The flux tubes are already studied  in great detail in $d=2+1$ dimensions for the  $SU(2)$ gauge group and some other groups (like $\mathcal{Z}_2$) and some studies are being conducted in this moment like with $SU(3)$ gauge group.

The intrinsic width of $k$-strings have been found \cite{Lucini:2001nv} to show little variations in $k$ for $SU(4)$ and $SU(5)$ in $d=3+1$ and for $SU(4)$ and $SU(6)$ in $d=2+1$.
The study of higher representations of $\mathcal{Z}_4$ in $d=2+1$ show the width of flux tubes is independent of  the specific representation of the source or of its $N$-ality and is universal.

One of the first studies demonstrating linear widening at finite temperature \cite{Allais:2008bk} has been performed in $d=2+1$ Ising model $\mathcal{Z}_2$. The finite temperature widening was also verified \cite{Gliozzi:2010zv} in detail with the $SU(2)$ group in $d=2+1$ dimensions. Moreover the connection between the zero temperature widening and the finite temperature widening was demonstrated  \cite{Gliozzi:2010jh}. Independence of the flux tube width on the lattice operators defining width was also demonstrated \cite{Caselle:2012rp}.

Moreover, widening in $SU(3)$ in $d=3+1$, the gauge group of QCD in the Standard Mode was also studied. In Ref. \cite{Bakry:2010sp}  the Ultraviolet filtering of QCD was explored at finite temperature, and the results exhibit a linearly divergent widening in agreement with the string picture predictions. At zero temperature,  Ref. \cite{Cardoso:2013lla} demonstrated logarithmic widening, and was also able to identify the intrinsic width, nonexistent in effective field theory. So far, the widening studies in $SU(3)$ have not yet applied the multilevel technique, resorting to simpler techniques to implement numerically such as smearing, multihit and groundstate amplification \cite{Cardoso:2013lla}, or Schwinger lines \cite{Cardaci:2010tb,Cea:2012qw}.

In what concerns the Abelian group $U(1)$, previous studies have not completely settled its flux tubes. 
In Ref. \cite{Zach:1997yz}, with the dual transformation technique, a logarithmic widening was found at $T=0$. 
In Ref. \cite{Koma:2003gi}, $d=3+1$ $U(1)$ was studied with the multilevel technique, however no clear signal of widening was found. 

The U(1) lattice gauge theory (also known as compact QED) is an abelian gauge theory whose group elements can be parametrized as a complex phase $U_\mu(x) \equiv \mbox{e}^{i \theta_\mu(x)}$. 
This gauge theory presents two phases separated by a phase transition \cite{Polyakov:1975rs}. In the first phase the theory is confining, featuring a string-like flux tube connecting any two quarks, while in the second one the flux tube disappears with the string tension going to zero. This phase is usually known as Coulomb phase due to including the normal electromagnetic Coulomb interaction in the weak coupling region.

Unlike the $\mathcal{Z}_N$ groups, the $U(1)$ group is continuous. It is easier to study in Lattice QCD than the $SU(N)$ continuous groups since it uses less computer memory. An interesting feature of $U(1)$ is that the electric and magnetic fields are gauge-invariant - unlike in the $SU(N)$ groups - and thus the electric and magnetic fields can be directly measured in lattice QCD with no gauge fixing. In   \cref{fig:field_charges_plane} we illustrate both two phases of the $U(1)$ theory, plotting our results for the electric field lines produced by a static charge and anticharge.

%
\begin{figure}[t!]
	\includegraphics[width=0.5\textwidth]{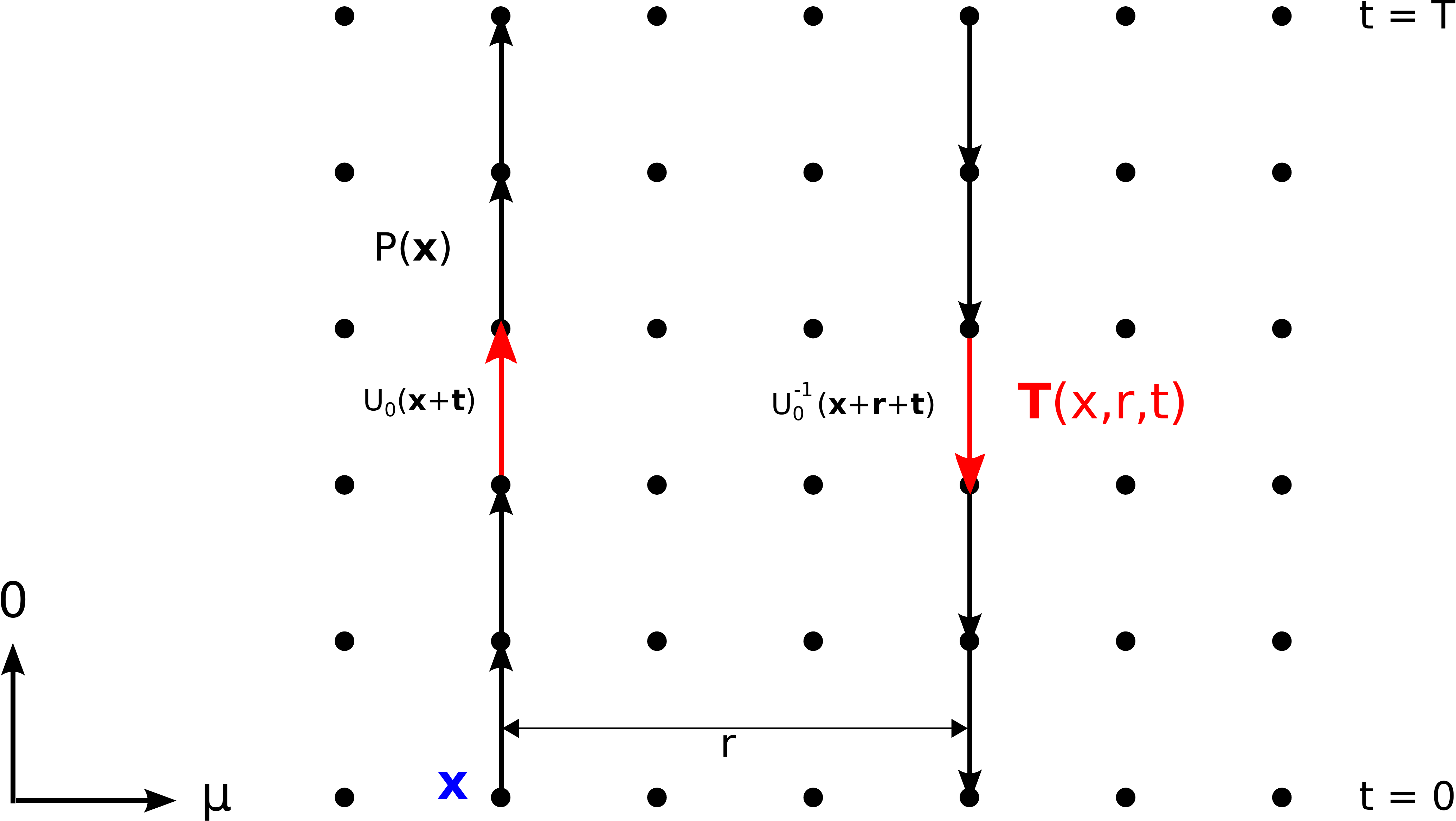}
	\caption{(color online) Two-link operator, factorized from the correlation of two Polyakov loops.}
	\label{fig:polyakov_two-link_operators}
\end{figure}

%
\begin{figure*}[ht!]
\begin{centering}
    \subfloat[Multilevel method for Polyakov loop correlations.]{
\begin{centering}
    \includegraphics[width=0.30\textwidth]{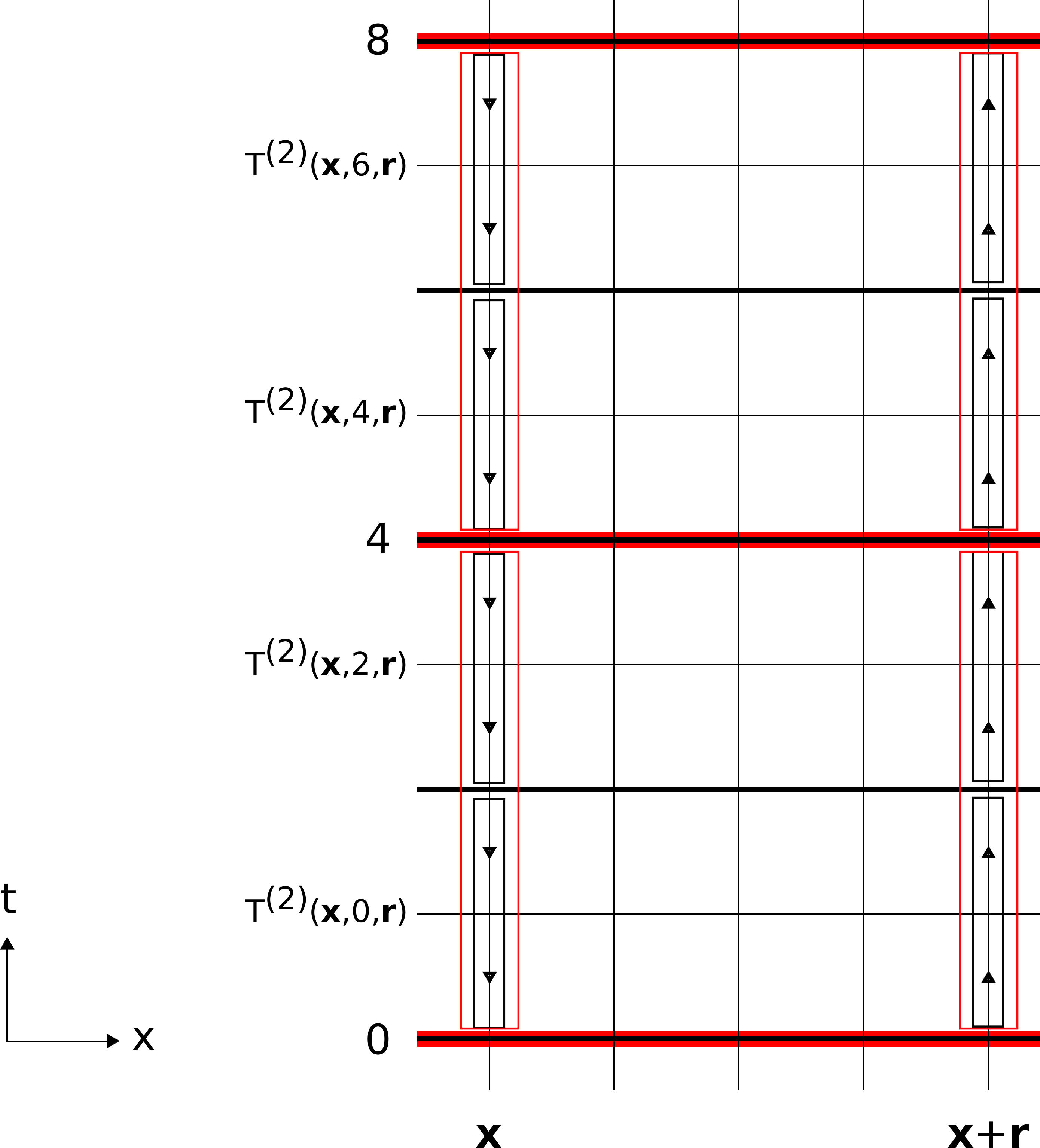}
\par\end{centering}}
    \subfloat[Multilevel method for field operators average.\label{el1}]{
\begin{centering}
    \includegraphics[width=0.60\textwidth]{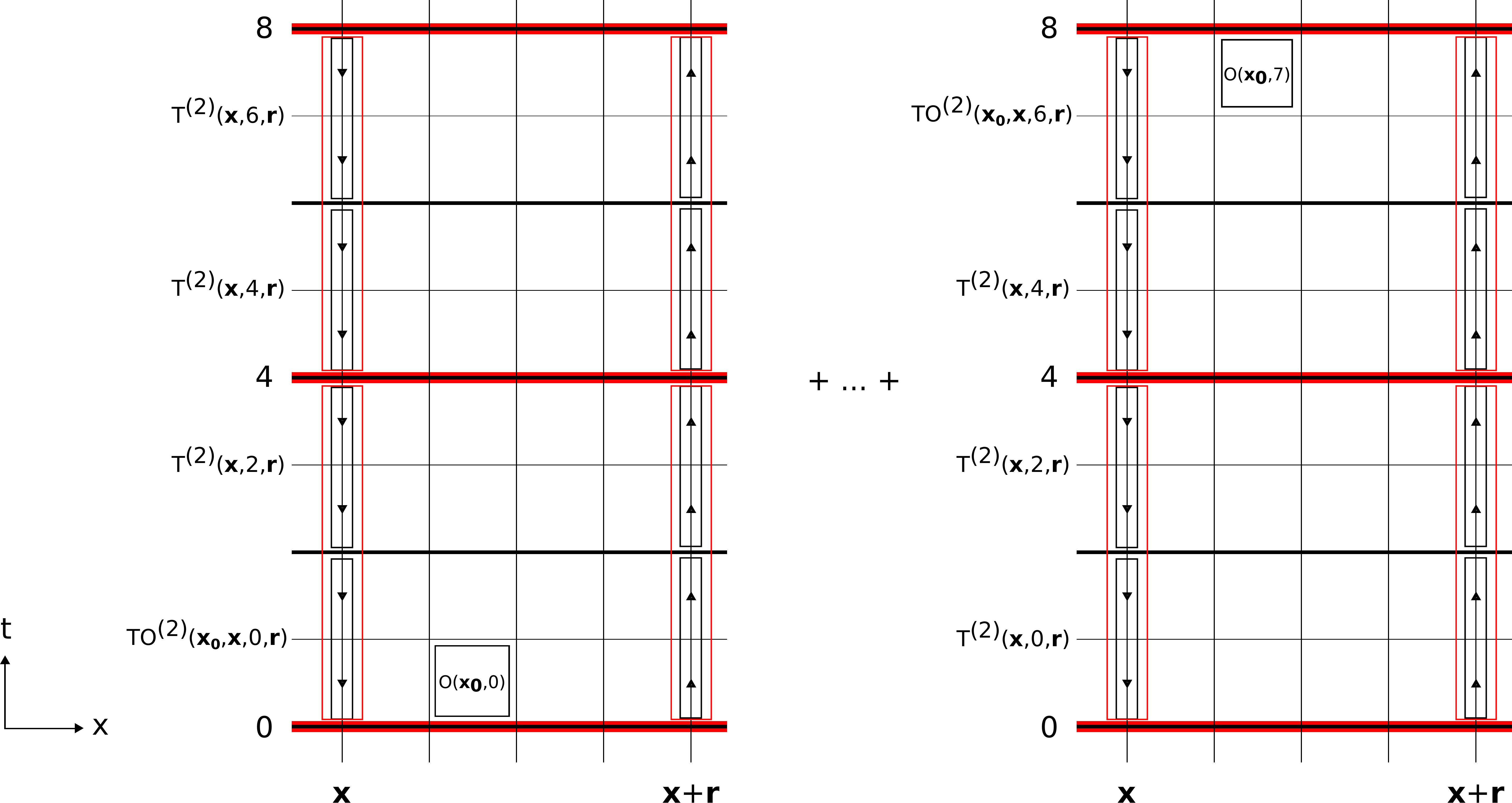}
\par\end{centering}}
\par\end{centering}
\caption{(color online) Multilevel method.}
\label{fig:multilevel_polyakov_correlations}
\end{figure*}

\subsection{Summary \label{sub:summary}}

In this work we focus on the numerical study of the string width for the $U(1)$ gauge group for different temporal extents. We aim to observe how the logarithmic widening at zero temperature changes into the linear widening of finite temperatures. We continue in the direction of  Ref. \cite{Koma:2003gi}, studying the widening of $d=3+1$ fluxtubes, at finite $T$, and utilizing the multilevel technique.

In this Section we motivate our work and introduce the results of Effective String Theory for the widening of the flux tube.  In  \cref{sec:methods} we define the numerical methods used in our calculations. To enhance the efficiency of our numerical computations, we write our codes in CUDA and run them on Graphics Processing Units (GPUs). We present the results in \cref{sec:results} . Finally a in \cref{sec:Conclusion} we summarize our results and present some paths worth exploring in the sequence of this work.


\section{Lattice setup and methods \label{sec:methods}}

We implement a lattice sized $24^3\times N_t$ with periodic boundary conditions. In this way, varying $N_t$ we obtain different time extents of the lattice in time direction thus different temperatures (we study $N_t = 4,\ 8,\ 12\ $ and $\ 24$).

The gauge group elements are parametrized as a complex phase $U_\mu(x) \equiv \mbox{e}^{i \theta_\mu(x)}$ and we use a standard Wilson action 
\begin{align}
S_g = \beta \sum_x \sum_{\mu<\nu} \mbox{Re}\left[1-P_{\mu\nu}(x)\right]
\end{align}
where $\beta$ is the inverse squared of the coupling constant $\frac{1}{g^2}$ and $P_{\mu\nu}(x) \equiv U_\mu(x)\ U_\nu(x+\hat{\mu})\ U^{-1}_\mu(x+\hat{\nu})\ U^{-1}_\nu(x)$ is the plaquette operator. Notice that, unlike SU(N) theories, no trace appears in the definition of the action due to the abelian nature of this theory.

For generating configurations we use the standard Metropolis algorithm combined with 3 steps of overrelaxation \cite{Creutz:PhysRevD.36.515,Bazavov:2009pj} per iteration. 

After lattice thermalization, we calculate the fields every 200 sweeps in order to ensure that the correlations between the various configurations we use are sufficiently low.


\subsection{Temperature
\label{sub:temperature}}

Temperature in lattice is defined as the inverse length of the lattice temporal extension, 
\begin{align}
T(N_t,\beta)=\frac{1}{N_t\ a(\beta)} \ ,
\end{align}
with $a(\beta)$ the lattice spacing.
The critical temperature $T_c$ above which the system is no longer in confining phase, is defined as
\begin{equation}
T_c=T(N_t,\beta_c) \ ,
\end{equation}
where $\beta_c$ is the critical $\beta$ (for our setup we determine $\beta_c \approx 1.003$).

For determining the temperature we need to introduce a scale in the lattice which allows us to estimate lattice spacing. We do this by calculating Sommer scaler $r_0$. This scale was introduced by Sommer in 1993 \cite{Sommer:1993ce} as an alternative to the string tension as scale parameter. He argues that this parameter has lower systematical and statistical errors than the traditional methods. 

It consists of choosing a specific point in the force, taken as the derivative of the potential, and using it as reference distance to normalize quantities. $r_0$ is defined through the relation $r_0^2F(r_0) = 1.65$.

We interpolate a function $F(r) = f_1 + \frac{f_2}{r^2}$ which constitutes a good local approximation to the force \cite{Sommer:1993ce}. To do so we compute the force $F(\bar{r}_r) = V(r)-V(r-1)$, as a discretized derivative of the potential and numerically solve the system
\begin{align}
f_1 + \frac{f_2}{\bar{r}^2_r} &= V(r)-V(r-1) \\
f_1 + \frac{f_2}{\bar{r}_{r+1}^2} &= V(r+1)-V(r) \\
r_0^2 \left(f_1 + \frac{f_2}{r_0^2} \right) &= 1.65
\end{align}
to obtain the $r_0$ value. The value of $\bar{r_r}$ is chosen to be a tree-level improved variable as to eliminate the $\mathcal{O}(a^2)$ term from $r_0$. This is done choosing the value of \cite{Sommer:1993ce}
\begin{align}
\bar{r_r} &= \left[ 4\pi \frac{G(r) - G(r-d)}{d} \right]^{-1/2}\\
G(r) &= \frac{1}{4 a} \int^\pi_{-\pi} \frac{\mbox{d}^3k}{(2 \pi)^3} \frac{\cos(k_1 r/a)}{4\sum_{j=1}^3 \sin^2(k_j/2)}.
\end{align}
This definition yields a considerable improvement for small $r_0$ over choice of $\bar{r}_r = r + \frac{a}{2}$, but becomes negligible at bigger values since the force becomes constant.

We use as an estimate to the error the difference of the result to the result including an extra term $1/r^4$ in the interpolation.

Having determined $r_0(\beta)=r_0/a(\beta)$ scale we can estimate the temperature at which we are working, since the temperature is $T=\frac{1}{N_t\ a(\beta)}$.
From here we can easily get
\begin{align}
\frac{T(\beta,N_t)}{T_c} = \frac{N_{tc}\ a(\beta_c)}{N_t\ a(\beta)} = \frac{N_{tc}\ r_0(\beta)}{N_t\ r_0(\beta_c)}
\end{align}
where $T_c$ is the reference critical temperature for confining phase transition. 

For determining $T_c$ we start by fitting the peak of plaquette susceptibility (which corresponds to the phase transition) in order to obtain $\beta_c$ for a lattice sized $24^3\times4$. Then we calculate $r_0(\beta_c)$ which allows us to compute $T_c = \frac{r(\beta_c)}{4 r_0}$.

%
\begin{figure*}
        \includegraphics[width=0.75\textwidth]{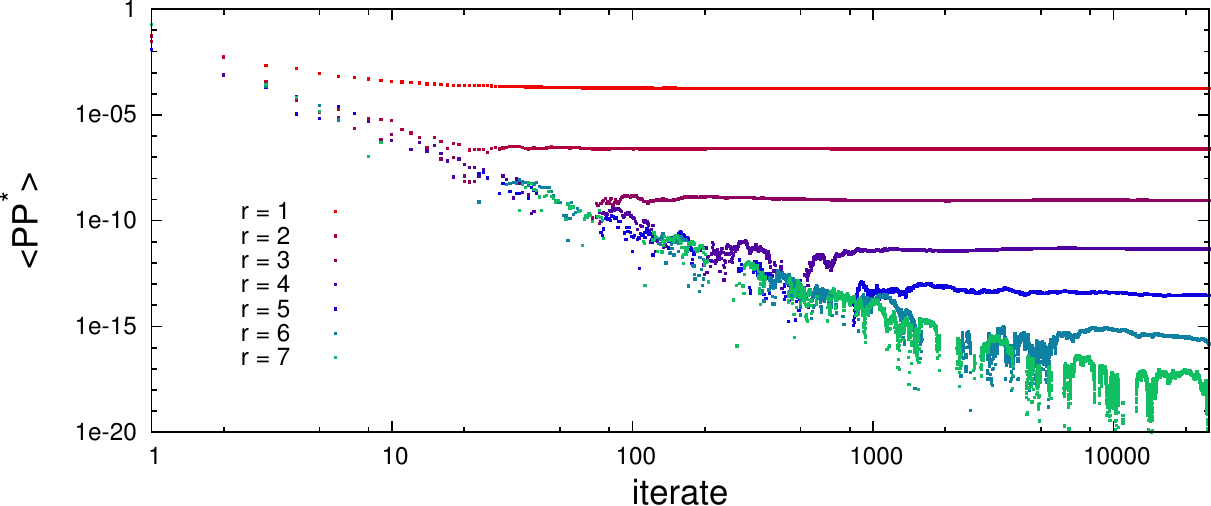}
    \caption{(color online) Multilevel convergence of the Polyakov loop average for different values of charge separation $r$ in a $24^4$ lattice with $\beta=1.00$. The scale is logarithmic in both axis.}
    \label{fig:multilevel_convergence}
\end{figure*}


\subsection{Charges and fields with \\ Polyakov loops and plaquetes 
\label{sub:polyakov}}

The operator we consider to place a static charge in the lattice is the Polyakov loop $P(\mathbf{x})$ 
\cite{Polyakov:1975rs,McLerran:1980pk}. 
A Polyakov loop is defined in lattice as the product of all the group elements in time direction in a specific point of space
\begin{align}
P(\mathbf{x}) \equiv \prod_{t=0}^{N_t-1} U_0(\mathbf{x}, t\hat{0})\ ,
\end{align}
where $U_0$ is the link along the temporal direction and $N_t$ the number of points the temporal direction.
This equation represents static charges in a certain point in lattice space.
The expectation value of a single Polyakov loop is given by
\begin{align}
\Braket{P} \equiv \Braket{\frac{1}{N_\sigma}\sum_\mathbf{x} P(\mathbf{x})}\ ,
\end{align}
where $N_\sigma = N_x\times N_y\times N_z$ is the lattice spatial volume.

The Polyakov loop average value \cite{McLerran:1980pk} follows $\Braket{P}\sim e^{a N_t\,F_Q}$, associated with the presence of a static, or infinitely heavy, charge with free energy $F_Q$ in the system. In the confining phase the average value of the Polyakov loop is vanishing, becoming non zero in the non confining-phase. As such it can be used as a parameter to study the phase transition between the confining and non confining phases.

The operator we choose to measure the fields in the lattice is the plaquette $P_{\mu\nu}(\mathbf x)$ . 
Expanding the plaquette as a function of the fields $F_{\mu\nu}(\mathbf x) $ we obtain,  
\begin{align}
P_{\mu\nu}(\mathbf x) &= \mbox{e}^{i a^2gF_{\mu\nu}+i\ensuremath{\mathcal{O}}(a^3)} = \nonumber \\
&= 1 + i a^2 g F_{\mu\nu} - \frac{a^4 g^2}{2}F^2_{\mu\nu} + i \ensuremath{\mathcal{O}}(a^3) + \ensuremath{\mathcal{O}}(a^5). \label{eq:plaq_expansion}
\end{align}
Separating the real and imaginary parts of this equation we obtain
\begin{align}
\mbox{Re} P_{\mu\nu}(\mathbf x) &= 1 - \frac{a^4 g^2}{2}F^2_{\mu\nu} + \ensuremath{\mathcal{O}}(a^5) \label{eq:real_part_plaq}\\
\mbox{Im} P_{\mu\nu}(\mathbf x) &= a^2 g F_{\mu\nu} + \ensuremath{\mathcal{O}}(a^3). \label{eq:imag_part_plaq}
\end{align}
and therefore the real part gives $F_{\mu\nu}^2(\mathbf x) = 1 - \mbox{Re} P_{\mu\nu}(\mathbf x) + \ensuremath{\mathcal{O}}(a^5)$. 
On the other hand, the imaginary part of the plaquette yields the value of $F_{\mu\nu}(\mathbf x)$.
Note that this imaginary term is not be accessible in non-Abelian groups because it would give $\mbox{Tr} F_{\mu\nu}(\mathbf x) = 0$. Therefore, in $U(1)$ we can measure both the expectation value of any component of the field directly and the expectation value of the square of any component.

At finite temperature, we can measure the squared field using the correlation between the Polyakov loop and the plaquette,
\begin{align}
\langle F_{\mu\nu}^2(\mathbf x) \rangle_{PP^\dagger} = \frac{\langle F_{\mu\nu}^2(\mathbf x) P P^\dagger \rangle}{\langle P P^\dagger \rangle} - \langle F_{\mu\nu}^2(\mathbf x) \rangle .
\end{align}
The width of the flux tube then translates to,
\begin{align}
w^2(r/2) = \frac{\int d^2x_\perp   x_\perp^2 \langle  F_{\mu\nu}^2(r,x_\perp) \rangle_{PP^\dagger}   }{\int d^2 x_\perp \langle  F_{\mu\nu}^2(r,x_\perp) \rangle_{PP^\dagger}   } \ ,
\label{eq:profile}
\end{align}
where $F_{\mu\nu}^2$ is extracted from a correlation between the Polyakov loop and the plaquette operator.


\subsection{Multilevel\label{multilevel}}

Some of the correlations that we need have a small average and a large variance, and thus a large relative error. For instance, in a $24^4$ lattice with $\beta=1.00$ and $r=6$ we the signal we want to measure is of the order of $\sim 10^{-15}$ for the Polyakov loop average, cf. \cref{fig:multilevel_convergence}. Hence we are forced to apply some method to reduce the relative errors faster than the $1/\sqrt{N_{cf}}$ we would have from pure Metropolis. 

We opt for implementing a L\"uscher and Wiese multilevel algorithm \cite{Luscher:2001up,Koma:2003gi} in order to reduce the number of configurations needed to obtain numerically stable results. 

Moreover, when calculating the Polyakov loop average we also apply the analytical multihit integration technique \cite{Parisi:1983hm,Koma:2003gi}.

The multilevel method was introduced by L\"uscher and Wiese in 2001 \cite{Luscher:2001up}, inspired in a generalization of multihit method.  It is useful for the calculation of the Polyakov or Wilson loops or other derivated quantities that involve products of links along the temporal direction (like the plaquette correlating with the Polyakov loop we use for measuring the field). 
Multilevel method explores the locality of the action by factorizing the path integral into smaller integrals calculated over sublattices. 



\subsubsection{Hierarchy in multilevel}

We now review the multilevel technique. We start by introducing two-link operators since they play an essential role in this formulation. Two-link operators are structures defined from the tensor product of two links in the same time layer at a certain distance $r$ (cf. figure \ref{fig:polyakov_two-link_operators}). This structure represents the propagation in time of a pair of fermions from time $t$ to time $t+1$ and they can be regarded as the basic constituents of the Polyakov loop correlation or of the temporal part of Wilson loop. They are defined as the following
\begin{align}
\Tone(\mathbf x, t, r\hat{\mu}) = U_0^\dagger (\mathbf x+t\hat{0})\ U_0(\mathbf x+r\hat{\mu}+t\hat{0}).
\end{align}

Notice that the two-link operator is a tensor in non-Abelian theories. 
In $U(1)$ the tensor product is simpler since it reduces to the product of two complex numbers.

Now we can rewrite the Polyakov loop correlation, in terms of these newly defined variables, as
\begin{align}
& \ P(\mathbf x)^\dagger P(\mathbf x+r\hat{\mu}) = \ \left(U_0(\mathbf x) \cdots U_0(\mathbf x+N_t\hat{0})\right)^\dagger 
\nonumber \\ & \hspace{2cm} \times
 \left(U_0(\mathbf x+r\hat{\mu}) \cdots U_0(\mathbf x+N_t\hat{0}+r\hat{\mu})\right) 
 \nonumber \\ 
 & \ \ \ \ \ = \Tone(\mathbf x, 0, r\hat{\mu}) \cdots  \Tone(\mathbf x, t, r\hat{\mu}) \cdots \Tone(\mathbf x, N_t, r\hat{\mu}).
\end{align}

The next step is to break this Polyakov loop correlation in a product of averages over sublattices.
To do so we need to find a way of isolating sublattices from within our lattice. 
These sublattices are time-slices of our original lattice, contained between two hyperplanes of constant time $x_0$ and $y_0$. 
If we take two of those hyperplanes and hold their spatial links constant we can isolate the dynamics of the sublattice from the rest of the lattice. 
This can be done thanks to the locality of the action (since the action on a link only depends on the plaquettes adjacent to it). This allows us to calculate subaverages of quantities inside the smaller lattices, sublattices. 
Following the usual convention, the sublattice expectation values are denoted by square brackets $[ \cdots ]$ and the expectation values over the whole lattice by $\langle \cdots \rangle$.

It is now possible to separate the integral in a hierarchical integration process with several intermediate levels. 
Since these integrals satisfy identities like $[\Tone(\mathbf x,t,r\hat{\mu})\Tone(\mathbf x,t+1,r\hat{\mu})] = [[\Tone(\mathbf x,t,r\hat{\mu})][\Tone(\mathbf x,t+1,r\hat{\mu})]]$, the Polyakov loop  correlation average can be calculated as
\begin{align}
\langle P^\dagger P \rangle =& \langle [[\Tone(\mathbf x,t,r\hat{\mu})][\Tone(\mathbf x,t+1,r\hat{\mu})]]
\nonumber \\
& \times [[\Tone(\mathbf x,t+2,r\hat{\mu})][\Tone(\mathbf x,t+3,r\hat{\mu})]] \cdots \nonumber \\
 \cdots &[[\Tone(\mathbf x,N_t-2,r\hat{\mu})][\Tone(\mathbf x,N_t-1,r\hat{\mu})]] \rangle.
\end{align}
It is easy to check that the innermost average corresponds to a multihit process because it fixes the spatial links everywhere and averages the link. 
In this way the multilevel algorithm can be seen as a generalization of multihit.


\alglanguage{pseudocode}
\algdef{SE}[LOOP]{LoopA}{EndLoopA}{\textbf{loop}}{\algorithmicend}
\begin{algorithm}[t!]
\begin{algorithmic}[1]
\hrule
\vspace{2pt}
\hrule
\vspace{3pt}
\LoopA \ $n_0$ times:
    \LoopA \ $k_0$ times:
        \State update whole lattice
    \EndLoopA
    \LoopA \ $n_4$ times:
        \LoopA \ $k_4$ times:
            \State update lattice freezing spacial links in 
            \State layers with $t$ multiple of 4
        \EndLoopA
        \LoopA \ $n_2$ times:
            \LoopA \ $k_2$ times:
                \State update lattice freezing spacial links in 
                \State layers with $t$ multiple of 2
            \EndLoopA
            \State Calculate $\Ttwo$
        \EndLoopA
        \State Calculate $[\Ttwo][\Ttwo]$
    \EndLoopA
    \State Calculate $P^\dagger P$
\EndLoopA
\State Calculate $\langle P^\dagger P\rangle$
\vspace{3pt}
\hrule
\vspace{2pt}
\hrule
\end{algorithmic}
\caption{Multilevel algorithm.
\label{alg:multilevel_alg}}
\end{algorithm}


\begin{table}[t!]
\begin{ruledtabular}
\begin{tabular}{ccc}
	~						& GeForce GTX 580 & Tesla C2070 \\ \hline
	CUDA capability 		&	2.0				&	2.0		\\ 
	Multiprocessors (MP)	&	16				&	14		\\ 
	Cores per MP			&	32				&	32		\\ 
	Total number of cores	&	512				&	448		\\ 
	Global memory			&	3072 MB GDDR5	&	6144 MB	GDDR5	\\
	Shared memory/SM		&	48 KB or 16 KB	&	48 KB or 16 KB	\\
	L1 cache/SM				&	16 KB or 48 KB	&	16 KB or 48 KB	\\
	L2 cache (chip wide)	&	768 KB			&	768 KB		\\
	Clock rate				&	1.57 GHz		&	1.15 GHz	\\ 
	Memory Bandwidth		&	192.4 GB/s		&	144 Gb/s	\\
	ECC support				&	no				&	yes		\\
	\end{tabular}
	\caption{Specifications of the used GPUs.}
	\label{tab:gpus}
	\end{ruledtabular}
\end{table}


\begin{figure*}[t]
        \includegraphics[width=0.8\textwidth]{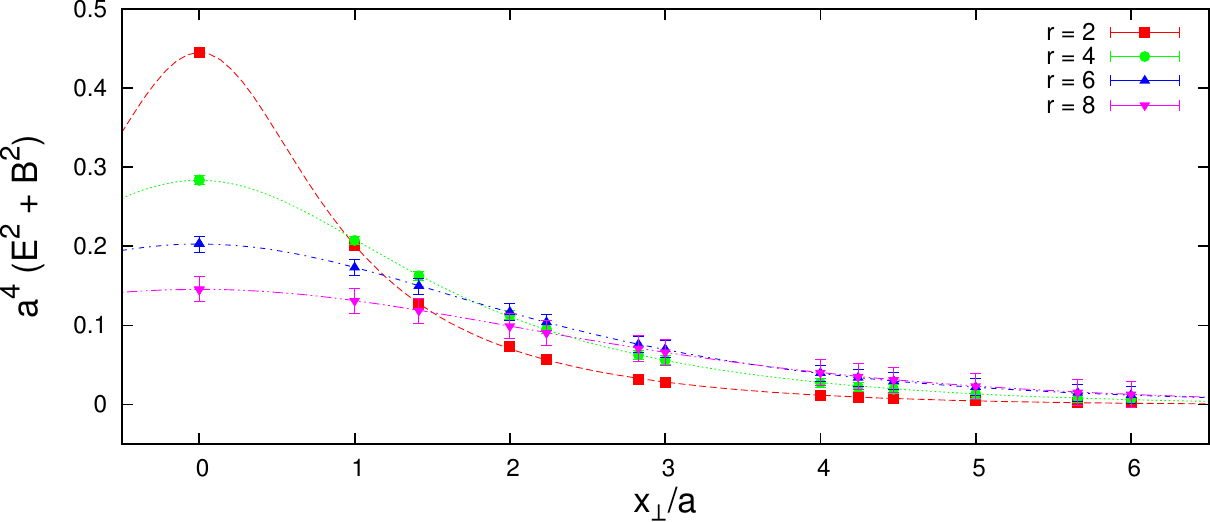}
    \caption{(color online) Flux tube profiles in the charges' mediator plan for $\beta=1$ and $N_t=4$ for several intercharge distances. }
    \label{fig:flux_tube_profile}
\end{figure*}


\subsubsection{Polyakov loop correlation}

To calculate the Polyakov loop correlation, it is useful to introduce auxiliary quantities that we average in the sublattices.
Here we follow approximately the notation defined in \cite{Koma:2003gi}. We start by defining the operator
\begin{align}
\Ttwo(\mathbf x,t,r\hat{\mu}) = \Tone(\mathbf x,t,r\hat{\mu}) \Tone(\mathbf x,t+1,r\hat{\mu}).
\end{align}
Defining the first average as an average in sublattices of thickness 2, we then have the Polyakov loop correlation,
\begin{align}
\langle P^\dagger P \rangle = &\langle [\Ttwo(\mathbf x,0,r\hat{\mu})][\Ttwo(\mathbf x,2,r\hat{\mu})]
\nonumber \\ & \cdots [\Ttwo(\mathbf x,N_t-2,r\hat{\mu})]\rangle.
\end{align}
In our code we implement one more level in order to achieve a further reduction on errors, calculating
\begin{align}
\langle P^\dagger P \rangle = &\langle [[\Ttwo(\mathbf x,0,r\hat{\mu})][\Ttwo(\mathbf x,2,r\hat{\mu})]] \dots \\  \nonumber
\cdots & [[\Ttwo(\mathbf x,N_t-4,r\hat{\mu})][\Ttwo(\mathbf x,N_t-2,r\hat{\mu})]]\rangle.
\end{align}

In practice the algorithm proceeds in a nested scheme as described in \cref{alg:multilevel_alg}.


\subsubsection{Electromagnetic Field}

From \cref{eq:real_part_plaq,eq:imag_part_plaq} we see how we can measure electromagnetic field tensor components, as well as its squares, from the plaquette operator,
\begin{align}
a^2 F_{\mu\nu} = \sqrt{\beta}\ \mbox{Im} P_{\mu\nu} + \ensuremath{\mathcal{O}}(a^3)
\end{align}
and
\begin{align}
a^4 F^2_{\mu\nu} = 2 \beta\ (1 - \mbox{Re} P_{\mu\nu}) + \ensuremath{\mathcal{O}}(a^6).
\end{align}

We want to study the field produced by static charges at a certain distance of each other so wecorrelate our operator with a Polyakov loop correlation representing those charges, getting
\begin{align}
\langle O \rangle_{P^\dagger P} = \frac{\langle P^\dagger P\, O \rangle}{\langle P^\dagger P \rangle} - \langle O \rangle
\end{align}
where $O$ stands for the operator we want to measure (either $F_{\mu\nu}$ or $F^2_{\mu\nu}$) and $\langle O \rangle_{P^\dagger P}$ stands for the expectation value of $O$ produced by the charges represented by $P^\dagger P$.

To implement a multilevel algorithm for this calculation it is useful to introduce some more quantities, besides $\Tone$ and $\Ttwo$,
\begin{align}
\TOtwo&(\mathbf x_0,\mathbf x,t,r\hat{\mu}) = [\Tone(\mathbf x,t,r\hat{\mu}) \Tone(\mathbf x,t+1,r\hat{\mu}) O(\mathbf x_0,t) + \nonumber \\
+& \Tone(\mathbf x,t,r\hat{\mu}) \Tone(\mathbf x,t+1,r\hat{\mu}) O(\mathbf x_0,t+1)]
\end{align}
\begin{align}
\Tfour(\mathbf x,t,r\hat{\mu}) = [\Ttwo(\mathbf x,t,r\hat{\mu})][\Ttwo(\mathbf x,t+2,r\hat{\mu})]
\end{align}
\begin{align}
\TOfour&(\mathbf x_0,\mathbf x,t,r\hat{\mu}) = \nonumber\\ 
=&\ [\Ttwo(\mathbf x,t,r\hat{\mu})][\TOtwo(\mathbf x_0,\mathbf x,t+1,r\hat{\mu})]\ + \nonumber \\
+&\ [\TOtwo(\mathbf x_0,\mathbf x,t,r\hat{\mu})][\Ttwo(\mathbf x,t+1,r\hat{\mu})].
\end{align}
With this notation we get,
\begin{align}
\langle & P^\dagger(0)P(r\hat{\mu}) O(x_0) \rangle = \frac{1}{{N_\sigma} N_t (d-1) }\sum_{\mathbf x}\sum_\mu 
\\ \nonumber
\phantom{+}
& [ \TOfour(\mathbf x_0,\mathbf x,0,r\hat{\mu})][\Tfour(\mathbf x,4,r\hat{\mu})] \cdots [\Tfour(\mathbf x,N_t-4,r\hat{\mu}) ]
\\ \nonumber
+& [ \Tfour(\mathbf x,0,r\hat{\mu})]  [ \TOfour(\mathbf x_0,\mathbf x,4,r\hat{\mu})] \cdots   [ \Tfour(\mathbf x,N_t-4,r\hat{\mu})] 
\\ \nonumber
+  & \ \ \cdots
\\ \nonumber
 +&  [ \Tfour(\mathbf x,0,r\hat{\mu})]  [ \Tfour(\mathbf x,4,r\hat{\mu})] \cdots  [ \TOfour(\mathbf x_0,\mathbf x,N_t-4,r\hat{\mu})]
\end{align}
and we compute it in a similar way to the Polyakov loop correlation.


\begin{figure*}
    \begin{tabular}{c c}
        \includegraphics[width=0.45\textwidth]{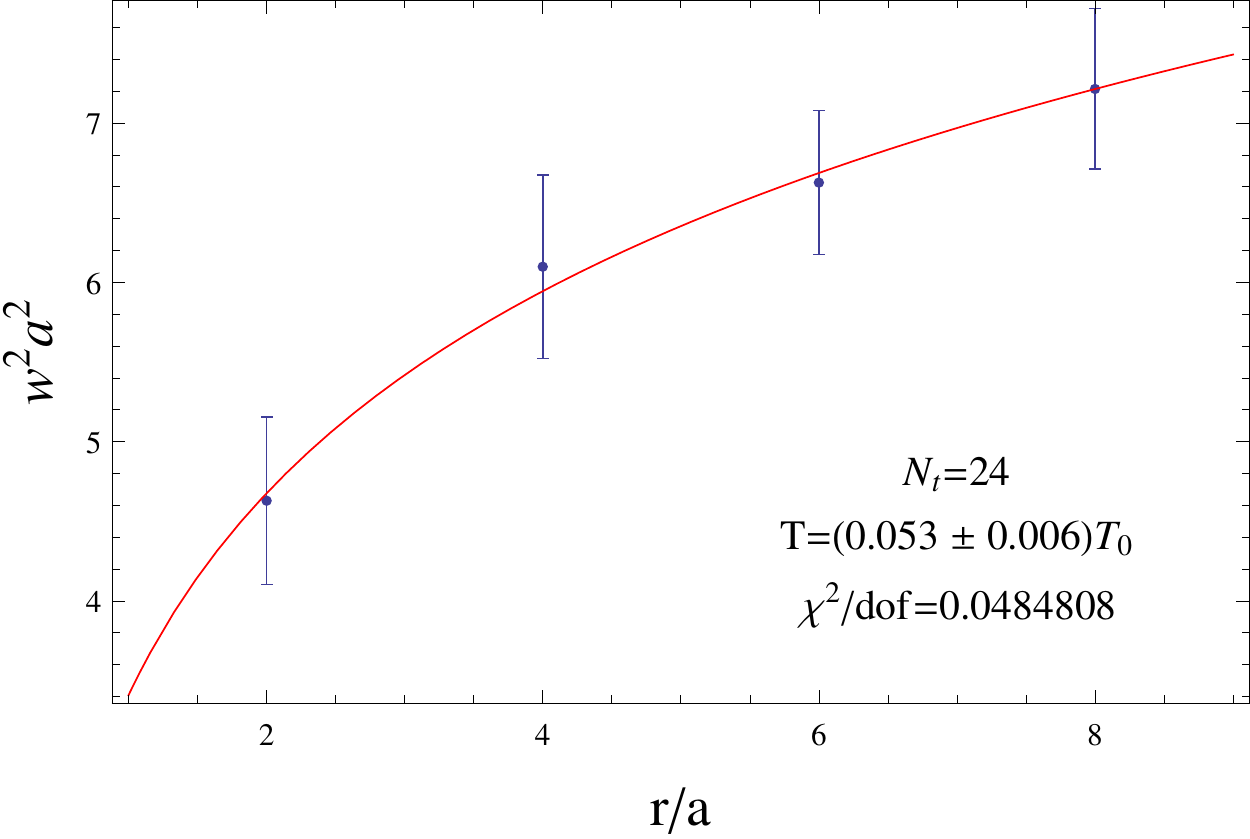} 
        &  \includegraphics[width=0.45\textwidth]{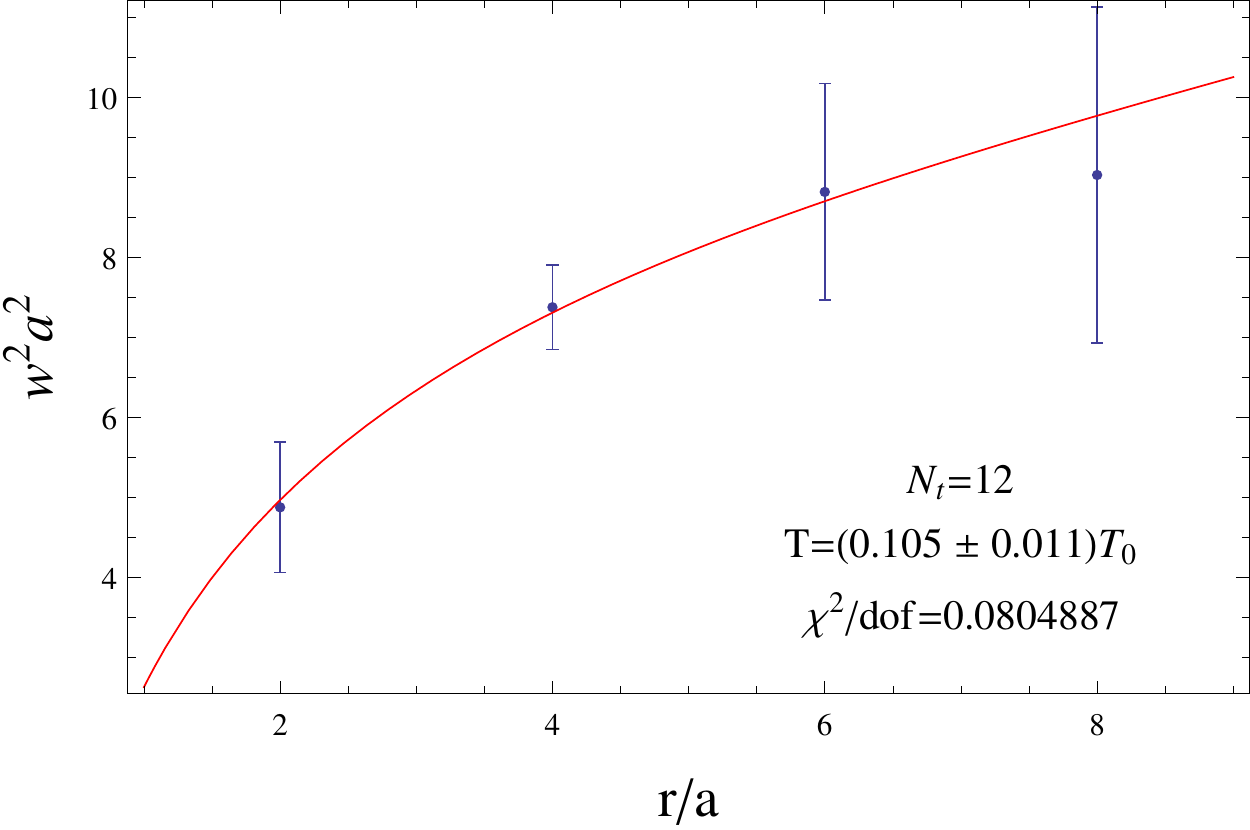} \\ 
        \includegraphics[width=0.45\textwidth]{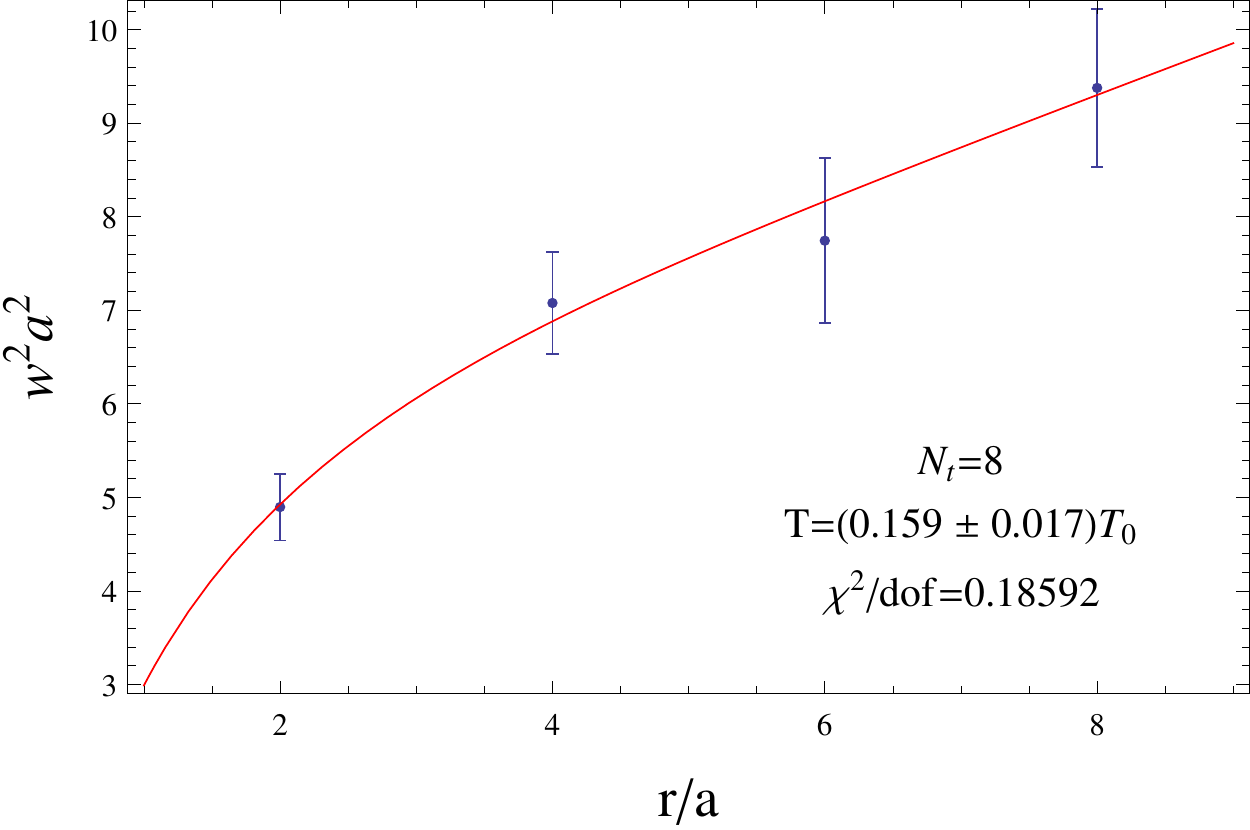}  
        &  \includegraphics[width=0.45\textwidth]{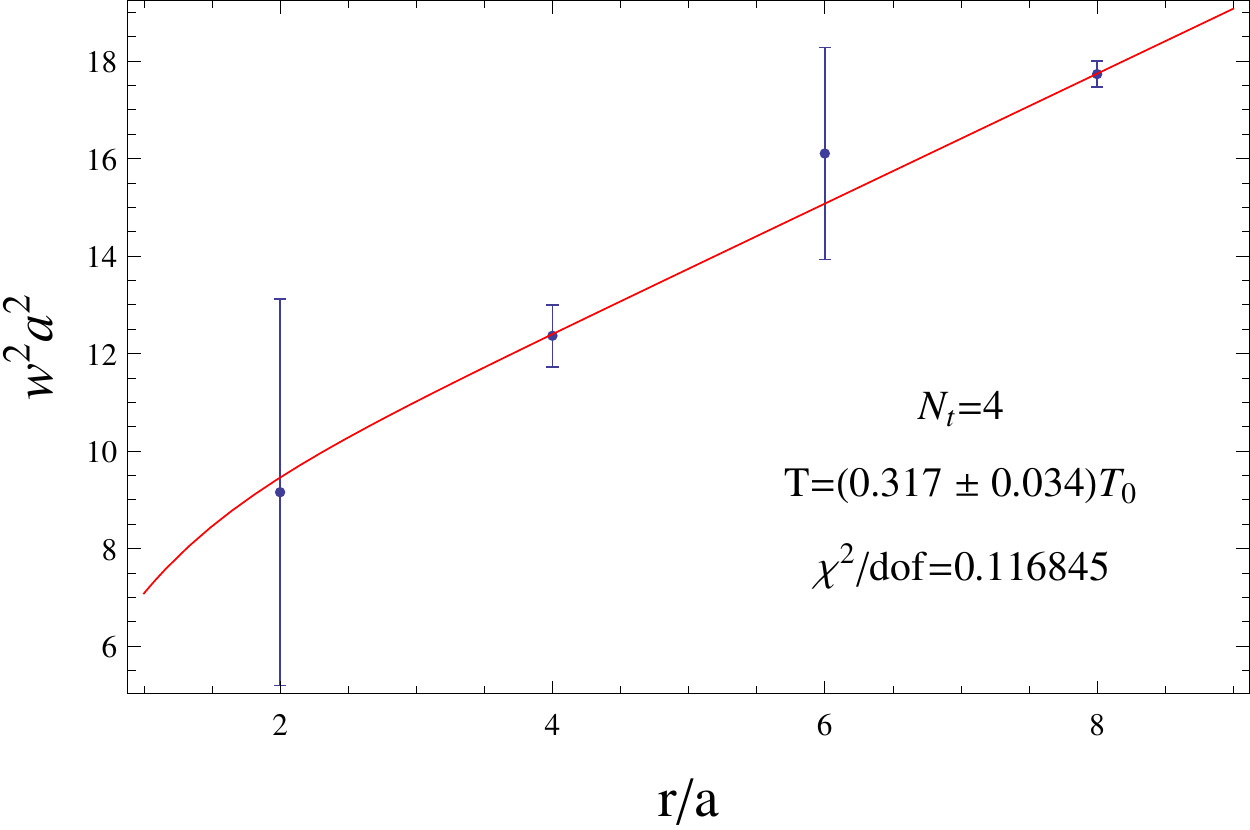}  
    \end{tabular}
    \caption{(color online) Squared width $a^2 w^2(r/a)$ for the different temperatures.}
    \label{fig:widths}
\end{figure*}


\subsection{Numerical Implementation in GPUs \label{sec:numerical_implementation}}

In our case, we notice in \cref{fig:multilevel_convergence} that the number of multilevel iterations needed to stabilize the result grows exponentially with the charge separation, since the correlations themselves fall exponentially with it. Moreover we want to work with large lattices and need many configurations to compensate high statistical errors. Thus we need to achieve high efficiency in our calculations.

As such we implement our code in CUDA and run it in GPUs,  where we can achieve high efficiency in parallel computations.
The GPUs we use are NVIDIA GTX 580 upgraded to 3 Gb memory, and NVIDIA Tesla C2070 with 6 Gb Memory, both FERMI generation GPUs (specifications in table \ref{tab:gpus}).

The GPU architecture is suited for lattice computations since it allows writing parallel kernels that can launch as many threads as the number of lattice sites, instead of utilizing cycles. This way we can take advantage of the high capacity of parallel computation of GPUs.

In our implementation we use the CUDA standard libraries, CUDA library CURAND, for the generation of random numbers, and the standard C++ libraries.

We work with double precision floating point numbers in all our computations. The $U(1)$ elements are parametrized by a phase which ensures that our links are always correctly projected on $U(1)$ manifold, except when we need to accumulate several links (e.g. for the calculation of averages), where we use a representation with two real numbers (for the real and imaginary parts). Double precision used to have a significantly lower performance compared to single precision in GPUs, but this fact has been considerably improved in the FERMI architecture.

We update the lattice in two steps: first the even points and then the odd points. This has the advantage of allowing us not to use a buffer lattice that would double the necessary memory on GPU effectively cutting the maximum size of the lattice we can run.

For the multilevel algorithm we need to accumulate results for the averages so we need to allocate extra lattices for that, making multilevel an algorithm that requires high availability of memory.

To measure the field with multilevel over a plan of the lattice we need to allocate a memory of $2\times$(size of double)$\times$(lattice volume)$\times$(area of the plan)/(multilevel step). For example, for a lattice with $N_t = 16$ and $N_\sigma = 24$ for a multilevel with step of $2$ we need to allocate an additional size of $2\times(8\text{ bytes})\times(24^3\times16)\times(24^2)/2 = 972 \text{MB}$, almost 600 times more than the $(8\text{ bytes})\times(24^3\times16) \approx 1.7 \text{MB}$ needed without multilevel for a lattice like this. Moreover in our code we utilize the global memory of the GPU, together with the cache memories L1 and L2.


\section{Results \label{sec:results}}


\subsection{Temperature dependence with $N_t$\label{sub:results_temperature}}

From the fit to the peak of plaquette susceptibility we obtain a value for critical $\beta$ in this lattice of $\beta_c \approx 1.003$. This corresponds to a value of $r_0/a = 9.10 \pm 0.97$, which yields $T_c \approx 2.3/r_0$.

Then we compute $r_0/a$ for $\beta = 1$ and $N_t = 48$, corresponding to $r_0/a$ at vanishing temperature, obtaining $r_0/a\,(1) = 2.8872\pm0.0002$. With this value we obtain the temperatures for different lattice temporal sizes in \cref{tab:widths}. The main source of error in the temperature is $r_0/a\,(\beta_c)$ determination since this is computed near the phase transition.

For $U(1)$ in compact QED there is no experimental data to compare with but if we take $r_0$ to have the same value as in QCD ($r_0 \approx 0.5 fm$ \cite{Sommer:1993ce}) we would find a value of $a \approx 0.2 fm$ for $\beta=1$.


\begin{table*}[t!]
\begin{ruledtabular}
\begin{tabular}{ccccccc}
$N_t$				& $T/T_c$ 							&$R$&$a^2 w^2(r/a)$  	&$a^2\sigma$						& $\rho_0/a$							&$\chi^2/dof$			\\ 	\hline
\multirow{4}{*}{24}	&\multirow{4}{*}{$0.053 \pm 0.006$}	& 2	&$4.62 \pm 0.53 $		&\multirow{4}{*}{$0.174 \pm 0.020$}	& \multirow{4}{*}{$0.155 \pm 0.052$}	& \multirow{4}{*}{$0.048$}	\\
					&									& 4	&$6.10 \pm 0.58$		&					&					& 				\\
					&									& 6	&$6.63 \pm 0.45$		&					&					& 				\\
					&									& 8	&$7.21 \pm 0.50$		&					&					& 				\\ \hline
\multirow{4}{*}{12}	&\multirow{4}{*}{$0.105 \pm 0.011$}	& 2	&$4.88 \pm 0.81$&\multirow{4}{*}{$0.094 \pm 0.017$}	& \multirow{4}{*}{$0.459 \pm 0.152$}	& \multirow{4}{*}{$0.080$}	\\
					&									& 4	&$7.38 \pm 0.53$&					&					& 				\\
					&									& 6	&$8.82 \pm 1.36$&					&					& 				\\
					&									& 8	&$9.03 \pm 2.10$&					&					& 				\\\hline
\multirow{4}{*}{8}		&\multirow{4}{*}{$0.159 \pm 0.017$}	& 2	&$4.90 \pm 0.36$&\multirow{4}{*}{$0.114 \pm 0.011$}	& \multirow{4}{*}{$0.343 \pm 0.078$}	& \multirow{4}{*}{$0.186$}	\\
					&									& 4	&$7.08 \pm 0.55$&					&					& 				\\
					&									& 6	&$7.74 \pm 0.88$&					&					& 				\\ 
					&									& 8	&$9.38 \pm 0.85$&					&					& 				\\ \hline
\multirow{4}{*}{4}		&\multirow{4}{*}{$0.317 \pm 0.034$}	& 2	&$9.16 \pm 3.96$&\multirow{4}{*}{$0.094 \pm 0.019$}	& \multirow{4}{*}{$0.124 \pm 0.140$}	& \multirow{4}{*}{$0.117$}	\\
					&									& 4	&$12.36 \pm 0.64$&					&					& 				\\
					&									& 6	&$16.10 \pm 2.17$&					&					& 				\\
					&									& 8	&$17.73 \pm 0.27$&					&					& 				\\ 
\end{tabular}
\caption{Temperature, widths, and parameters fitting the widening with the effective string theory prediction.}
\label{tab:widths}
\end{ruledtabular}
\end{table*}


\subsection{Flux tube profile\label{sub:results_fluxtube_profile}}

As an illustration of the technique we use to calculate the field, we show figures of the electric field in the plane of the charges in confining and non-confining (Coulomb) phase (cf. \cref{fig:field_charges_plane}). 

Applying the techniques described in \cref{sec:methods} we obtain the profile of $\langle F_{\mu\nu}^2(x) \rangle_{PP^\dagger}$  in the mediator plane of the charge and anti-charge. Considering all components of the fields, we compute the expectation value of the energy density $\langle E^2 + B^2 \rangle_{PP^\dagger}$. We consider different $N_t = 4,\ 8,\ 12\ $ and $\ 24$ in order to obtain a significant range of temperatures. 

The mediator plane profiles are necessary for the computation of the width of the flux tube. An example of the profiles we obtain is in  \cref{fig:flux_tube_profile}. In this example we generate 100 multilevel configurations.


\subsection{Width determination and comparison with theoretical results\label{sub:width_determination}}

To this profile we fit the Ansatz suggested in \cite{Gliozzi:2010jh}
\begin{align}
\frac{\langle P^\dagger P\ F_{\mu\nu} \rangle}{\langle P^\dagger P \rangle} = C_1 \exp (-x_\perp^2/s)\frac{1+C_2\exp (-x_\perp^2/s)}{1+C_3\exp (-x_\perp^2/s)} \
\end{align}
in order to interpolate between the points we measure. Fitting the results for the profiles with \cref{eq:profile},  and integrating them
we calculate the square widths of the flux tubes. 
To obtain an error estimation of the square width w, $w^2(r/a)$, we use the jackknife method. Our results are presented in 
 \cref{tab:widths} and \cref{fig:widths}.


\subsection{Width of the flux tube for different temperatures\label{sub:results_width}}

To the results of the width, we fit the theoretical expression \cref{eq:width_tree_level} from effective string theory, as a  function of the intercharge distance. Again we apply the jackknife method for the estimation of the errors of our fit parameters.

Clearly, the flux tube width consistently grows with the charge separation. It is also interesting to observe in  \cref{fig:widths} how the logarithmic behavior at low temperature shifts to the high temperature linear behavior.

With $N_t = 24$, close to zero temperature regime, we obtain the expected almost pure logarithmic behavior. When decreasing the temporal extent of the lattice, thus increasing the temperature, we progressively start observing deviations from pure logarithmic behavior, tending to the linear high temperature limit. With $N_t = 4$, the highest temperature studied, the linear behavior of the curve already sets in most of the range of studied distances.

In what concerns the parameters $\sigma$ and $\rho_0$ of our ansatz in  \cref{eq:width_tree_level}, they fluctuate with $T$. Although with fluctuations slightly larger than the error bars,  they may be compatible with  being constant in the considered temperature range $T \in [0,  T_c /3]$. Moreover, the string tension $\sigma$ is comparable to the string tension $\sigma = 0.180 \pm 0.013$ we obtain fitting \cref{eq:potential} with the charge-anticharge potential computed on the lattice. Thus we find reasonable agreement with the effective string theory.


\section{Conclusion and foreword
\label{sec:Conclusion}}

We compute the width $w$ of the $U(1)$ flux tube in the mediator plane of the charge and anticharge, for different temperatures and different intercharge distances, and exhibit our results in \cref{fig:widths,fig:flux_tube_profile,fig:multilevel_polyakov_correlations,tab:widths}.

It occurs that the $U(1)$ group is not trivial to study although it is Abelian. We find the fluctuations are surprisingly large in the confining sector. The non-confining Coulomb phase is quite different from the confining phase, since its numerical results are quite stable. Therefore the confining phase requires a large number of configurations, and a very efficient numerical noise suppression technique. We apply the multilevel method, and to maximize the efficiency or our numerical calculations, we run our lattice gauge theory computations in GPUs.

Clearly, the flux tube width consistently grows with the charge separation. It is also interesting to observe in  \cref{fig:widths} how the logarithmic behavior at low temperature shifts to the high temperature linear behavior.  The widths are fitted with the effective string theory, with a small $\chi^2/dof$. We find reasonable agreement for constant string tension $\sigma$  and distance parameter $\rho_0$ in the studied temperature range $T \in [0,  T_c /3]$.

The quality of our results suggest the present study may be extended to further understand the details of the compact $U(1)$ flux tube, beyond the effective string theory. It would be very interesting to compare $\langle F_{\mu\nu}^2(x) \rangle_{PP^\dagger}$ with $\langle F_{\mu\nu}(x) \rangle_{PP^\dagger}^2$,  to study the widening close to $T_c$ where the string tension $\sigma$ is expected to decrease,  or to extract the intrinsic width of the flux tube. Moreover confinement in $U(1)$ can also be applied, possibly coupling the electromagnetic field with a scalar Higgs field, to superconductivity. 

\begin{acknowledgments}
This work was partly funded by the FCT contracts,  POCI/FP/81933/2007, 
CERN/FP/83582/2008, PTDC/FIS/100968/2008, CERN/FP/109327/2009, CERN/FP/116383/2010 and CERN/FP/123612/2011.
\end{acknowledgments}

\bibliographystyle{apsrev4-1}
\bibliography{bib}

\end{document}